**Electronic structure of the 344-type superconductors La$_3$(Ni;Pd)$_4$(Si;Ge)$_4$ by *ab initio* calculations**

M.J. Winiarski[1] and M. Samsel-Czekała[1] *

[1] *Institute of Low Temperature and Structure Research, Polish Academy of Sciences, P.O. Box 1410, 50-950 Wrocław 2, Poland*

**Abstract**

Electronic structures of superconducting ternaries: La$_3$Ni$_4$Si$_4$, La$_3$Ni$_4$Ge$_4$, La$_3$Pd$_4$Si$_4$, La$_3$Pd$_4$Ge$_4$, and their non-superconducting counterpart, La$_3$Rh$_4$Ge$_4$, have been calculated employing the full-potential local-orbital method within the density functional theory. Our investigations were focused particularly on densities of states (DOSs) at the Fermi level with respect to previous experimental heat capacity data, and Fermi surfaces (FSs) being very similar for all considered here compounds. In each of these systems, the FS originating from several bands contains both holelike and electronlike sheets possessing different dimensionality, in particular quasi-two-dimensional cylinders with nesting properties. A comparative analysis of the DOSs and FSs in these 344-type systems as well as in nickel (oxy)pnictide and borocarbide superconductors indicates rather similar phonon mechanism of their superconductivity.

**1. Introduction**

Intermetallics forming layered PbO-type structures have recently drawn wide interest since the discovery in 2008 of high-temperature (high-$T_C$) superconducting iron (oxy)pnictides reaching $T_C$ up to 55 K [1]. However, isostructural with them nickel-based (oxy)pnictides do not exhibit such high $T_C$'s [2,3,4]. The pnictidelike structures are strongly anisotropic (quasi-two-dimensional, Q2D), being built from positively charged layers of atoms of alkaline or rare-earth metals and negatively charged layers containing transition metals and non-metallic atoms. The comparison of electronic structure between similar superconducting systems but based on various transition-metal elements may be crucial in understanding mechanisms of high-$T_C$ SC in the iron-based class of these compounds and its lack in other ones.

Four superconductors: La$_3$Ni$_4$Si$_4$ ($T_C$ ~ 1.0 K), La$_3$Ni$_4$Ge$_4$ ($T_C$ ~ 0.8 K), La$_3$Pd$_4$Si$_4$ ($T_C$ ~ 2.2 K), and La$_3$Pd$_4$Ge$_4$ ($T_C$ ~ 2.8 K) have been reported [5-9] among orthorhombic 344-type systems. They adopt the U$_3$Ni$_4$Si$_4$-type structure (*Immm*), resembling that of (oxy)pnictides, with a large ratio ($c/a$ ~ 6). Among this kind of ternaries, non-superconducting representatives are La$_3$Rh$_4$Ge$_4$ [10], Ce$_3$Rh$_4$Ge$_4$, Ce$_3$Rh$_3$IrGe$_4$ [11] and Kondo-lattice Ce$_3$Pd$_4$Ge$_4$ [12] systems. On the other hand, lanthanides in the $Ln_3$Pd$_4$Ge$_4$ family ($Ln$ = Y, Gd, Tb, Dy, Ho, Er, Tm and Yb) [13-18] with other orthorhombic structure of the Gd$_3$Cu$_4$Ge$_4$-type, forming Pd-Ge cages, exhibit no superconductivity (SC). A comparison between the structural properties of the orthorhombic La$_3$Pd$_4$Ge$_4$ and tetragonal LaPd$_2$Ge$_2$ (T$_C$ ~ 1.1 K) systems suggested that SC in this class of compounds is sensitive to an arrangement of transition-metal and non-metal atoms as in the Pd-Ge network [7].

In this work, we study electronic structure of the lanthanum 344-type family of superconductors: La$_3$Ni$_4$Si$_4$, La$_3$Ni$_4$Ge$_4$, La$_3$Pd$_4$Si$_4$, La$_3$Pd$_4$Ge$_4$, and their non-superconducting reference compound, La$_3$Rh$_4$Ge$_4$, by *ab initio* methods. Our investigations are focused on a relation between their densities of states (DOSs) at the Fermi level (E$_F$) and T$_C$ values. Also their Fermi surfaces (FSs) topology is compared with that of La$_3$Rh$_4$Ge$_4$. Furthermore, we discuss their electronic structure similarities to those of (oxy)pnictide and borocarbide superconductors.

## 2. Computational methods

Electronic structure of all considered here 344-type compounds, crystallizing in the orthorhombic *Immm* structure (visualized in Fig. 1), have been computed with the full-potential local-orbital (FPLO-9) method [19]. The Perdew-Wang form [20] of the local density approximation (LDA) of exchange-correlation functional was employed in the scalar relativistic mode.

X-ray diffraction values of lattice parameters (in nm) used in our calculations, taken from Ref. [7], are as follows: *a* = 0.41305, *b* = 0.41760, *c* = 2.3578 for La$_3$Ni$_4$Si$_4$; *a* = 0.42017, *b* = 0.42167, *c* = 2.4031 for La$_3$Ni$_4$Ge$_4$; *a* = 0.42254, *b* =0.43871, *c* = 2.4551 for La$_3$Pd$_4$Si$_4$; *a* = 0.42293, *b* = 0.43823, *c* = 2.50109 for La$_3$Pd$_4$Ge$_4$; *a* = 0.41746, *b* = 0.42412, *c* =2.5234 for La$_3$Rh$_4$Ge$_4$. The experimental atomic positions, given in [7], were assumed as initial ones and then optimized for all investigated compounds to minimize forces in the unit cell (u.c.), containing double formula unit (f.u.), prior to further band structure computations. Valence-basis sets were automatically selected by the FPLO-9 internal procedure. Total energy values of the considered here systems were converged with accuracy to ~1 meV for the 16x16x16 *k*-point mesh, yielding 621 *k*-points in the irreducible part of the Brilouin zone (BZ).

## 3. Results and discussion

The total and partial DOSs of five probed here systems are plotted in Fig. 2. Their overall shapes, being similar to one another, are dominated by the broad peaks of the transition-metal (Ni, Pd, Rh) 3d/4d and non-metal (Si,Ge) 3p/4p, as well as La 5d electron orbitals. The highest peaks of the Ni 3d and Rh 4d electrons are centered at about 1.5-2.0 eV below the Fermi level while those coming of a greater number of Pd 4d electrons are located deeper in energy, i.e. 3.0-3.5 eV below E$_F$. Hence, they have slightly smaller contributions at E$_F$ than those of the Ni-based systems. For the considered here ternaries, the total DOSs at E$_F$, N(E$_F$), are collected in Table 1. It presents both our calculated (LDA) and experimental N(E$_F$), derived from heat capacity data of Ref. [9]. This table indicates that the LDA values of N(E$_F$) are all slightly smaller than the corresponding experimental data. Nevertheless, both theoretical and experimental N(E$_F$) are inversely proportional to the superconducting transition temperatures, $T_C$'s. This fact can be explained by the strong influence of the electron-phonon coupling on SC in this family of compounds as postulated in Ref. [9].

As is visible in Fig. 2, in La$_3$Ni$_4$(Si;Ge)$_4$ the Fermi level is placed near the narrow sub-peak originating mainly from the Ni 3d electrons, which resembles the situation taking place in other nickel-based systems possessing the *I4/mmm* symmetry [21-22]. It enables further $T_C$ tuning by any change in DOS at $E_F$, in analogy to the borocarbides [23], where the Fermi level is situated exactly at the local maximum of DOSs. For the stronger electron–phonon coupled palladium-based systems [9], their DOS's in the vicinity of $E_F$ are diminished and become quite flat. In consequence, one may expect rather insignificant pressure effects on the electronic structure and related properties of this series.

Our calculated electronic occupation numbers, $N_{calc}$, in the investigated here 344-type compounds, compared with those for the free atoms, $N_{at}$, are given in Table 2. The electron populations for La, Si and Ge orbitals are similar in all considered here systems. Namely, for La 5d electrons $N_{calc}$ = 1.4, being enhanced compared with $N_{at}$= 1. This is in contrary to strongly reduced $N_{calc}$ (~ 0.2) of the La 6s electrons with respect to its $N_{at}$ (= 2). In turn, $N_{calc}$'s for both the Si 3s and Ge 4s electrons have slightly diminished $N_{calc}$ (~ 1.5) in relation to their $N_{at}$ (= 2). Whereas $N_{calc}$'s for the Si 3p and Ge 4p electrons are substantially higher (~ 3.0) than $N_{at}$ (= 2). Interestingly, in both Ni-based superconductors, the Ni 3d electrons have $N_{calc}$ = 8.8, being strongly enhanced with respect to $N_{at}$ = 8. This is opposite behavior to that of the Pd-based superconductors, where the Pd 4d electrons possess much reduced $N_{calc}$ (~ 9.0) compared with $N_{at}$ (= 10). Meanwhile, in the case of the non-superconducting compound with rhodium, for the Rh 4d electrons, $N_{calc}$ ~ 8.3 is only slightly increased ($N_{at}$ = 8). Finally, Ni/Pd/Rh 4s/5s and 4p/5p electrons have similar $N_{calc}$ ~ 0.5, though they have different values of $N_{at}$.

The calculated bands energies, $\varepsilon_n(\mathbf{k})$, are plotted in Fig. 3 only for two chosen superconductors, i.e. La$_3$Ni$_4$Si$_4$ and La$_3$Pd$_4$Ge$_4$, and also for the reference compound, La$_3$Rh$_4$Ge$_4$. As this figure indicates, the band structures of both mentioned above superconductors as well as of the other ones (not displayed here), considered in this paper, are quite similar to one another at $E_F$, compared with a reduced number of bands cutting $E_F$ in the non-superconducting La$_3$Rh$_4$Ge$_4$. It is also seen in Fig. 3 that the band structure in these superconductors could be the most sensitive to pressure or doping along the ΓX and XU1 lines.

The Fermi surfaces of the considered above three compounds, i.e. La$_3$Ni$_4$Si$_4$, La$_3$Pd$_4$Ge$_4$ and La$_3$Rh$_4$Ge$_4$, are displayed in Fig. 4. As is visible in this figure, the FSs are quite similar for the nickel- and palladium-based superconductors, originating from as many as six bands, while in the reference rhodium-based system its FS comes from only five bands. All FSs contain three-dimensional (3D) holelike pockets and complex sheets of I and II conduction bands. In turn, the electronlike FS sheets in III-VI bands form Q2D corrugated cylinders, centered at the S points, being also characteristic of other pnictidelike superconductors [21,22,24]. Such complex FSs allow for multi-band SC, particularly as in the case of borocarbides [23,24]. It also turned out that in the nikel-borocarbide LuNi$_2$B$_2$C superconductor, specific nesting features in electronlike FS sheets can be responsible for strong Kohn (electron-phonon) anomalies, enhancing rather BCS-like SC [24]. It should be underlined here that only for the palladium-based superconductors, we have obtained, in our LDA calculations, an analogous perfect nesting with the vector **q** = 0.5 x (2π/**b**), which is drawn in Fig. 5. It is also seen in this figure that such a nesting is much less perfect in the case of the nickel-based superconductors but completely lacking in the non-superconducting $La_3Rh_4Ge_4$ system. This nesting feature might explain an enhancement of $T_C$ for the palladium-based superconductors as connected with possible electron-phonon anomalies, which requires additional experimental studies.

## 4. Conclusions

The band structures of four superconductors $La_3(Ni;Pd)_4(Si;Ge)_4$ and the non-superconducting reference system, $La_3Rh_4Ge_4$, have been studied from the first principles. Our calculated densities of states at the Fermi level are inversely proportional to the $T_C$'s, which supports the assumption, derived from previous experimental data, that superconductivity in this class of compounds can be strongly driven by the strength of the electron–phonon coupling. Therefore, as expected, higher $T_C$ are reached by systems with lower DOSs at $E_F$. The multi-band Fermi surfaces of all investigated here superconductors, allowing for an occurrence of multi-gap superconductivity, are similar to one another. The FSs originate from six bands and contain three-dimensional holelike pockets and complex sheets as well as quasi-two-dimensional corrugated electronlike cylinders. The nesting properties of electron III FS sheet of $La_3Pd_4Ge_4$ indicates that superconductivity for palladium-based compounds may be additionally enhanced due to electron-phonon anomalies, as it happens in the case of nickel borocarbide superconductors.

**Acknowledgments**

The National Center for Science in Poland is acknowledged for financial support of Project No. N N202 239540. The calculations were carried out mainly in Wroclaw Center for Networking and Supercomputing (Project No. 158). The Computing Center at the Institute of Low Temperature and Structure Research PAS in Wrocław is also acknowledged for the use of the supercomputers and technical support.

**Tables:**

**Table 1** Our computed (LDA) total DOSs at $E_F$, $N(E_F)$ (in electrons/eV/f.u.) and available experimental values of $N(E_F)$ and maximum $T_C$'s (in K), for the $La_3(Ni;Pd)_4(Si;Ge)_4$ ternaries [5-9].

| compound | calculated $N(E_F)$ | experimental $N(E_F)$ | maximum $T_C$ |
|---|---|---|---|
| $La_3Ni_4Si_4$ | 5.1 | 5.3 – 5.4 [9] | 1.0 [8,9] |
| $La_3Ni_4Ge_4$ | 5.0 | 5.0 – 5.2 [9] | 0.8 [9] |
| $La_3Pd_4Si_4$ | 4.2 | - | 2.2 [6] |
| $La_3Pd_4Ge_4$ | 3.6 | 3.7 – 3.9 [9] | 2.8 [5,7] |
| $La_3Rh_4Ge_4$ | 3.4 | - | - |

**Table 2** Calculated (LDA) electron occupation numbers, $N_{calc}$, (per given orbital at single atomic position) for $La_3T_4(Si;Ge)_4$ systems (with accuracy to ± 0.1), compared with the corresponding numbers for free atoms, $N_{at}$ (given in parentheses). It should be noticed that in each system, $N_{calc}$ of the same electron orbitals in $T$ and Si/Ge atoms are varying (± 0.4) depending on their atomic positions in the u.c., from which given orbitals originate.

| compound | La 5d | La 6s | $T$ 3d/4d | $T$ 4s/5s | $T$ 4p/5p | Si 3s/ Ge 4s | Si 3p/ Ge 4p | Si 3d/ Ge 4d |
|---|---|---|---|---|---|---|---|---|
| $La_3Ni_4Si_4$ | 1.4 (1) | 0.2 (2) | Ni 3d 8.8 (8) | Ni 4s 0.6-0.8 (2) | Ni 4p 0.5 (0) | 1.4-1.5 (2) | 2.8-2.9 (2) | 0.3 (0) |
| $La_3Ni_4Ge_4$ | | | | | | 1.5-1.6 (2) | 2.8-3.2 (2) | 0.1 (0) |
| $La_3Pd_4Si_4$ | | | Pd 4d 9.0 (10) | Pd 5s 0.6-0.7 (0) | Pd 5p 0.5 (0) | 1.5 (2) | 2.8-3.0 (2) | 0.2-0.3 (0) |
| $La_3Pd_4Ge_4$ | | | Pd 4d 9.0-9.1 (10) | | Pd 5p 0.4 (0) | 1.6 (2) | 2.9-3.2 (2) | 0.1-0.2 (0) |
| $La_3Rh_4Ge_4$ | | | Rh 4d 8.2-8.4 (8) | Rh 5s 0.5-0.6 (1) | Rh 5p 0.3 (0) | 1.6-1.7 (2) | 2.9-3.1 (2) | |

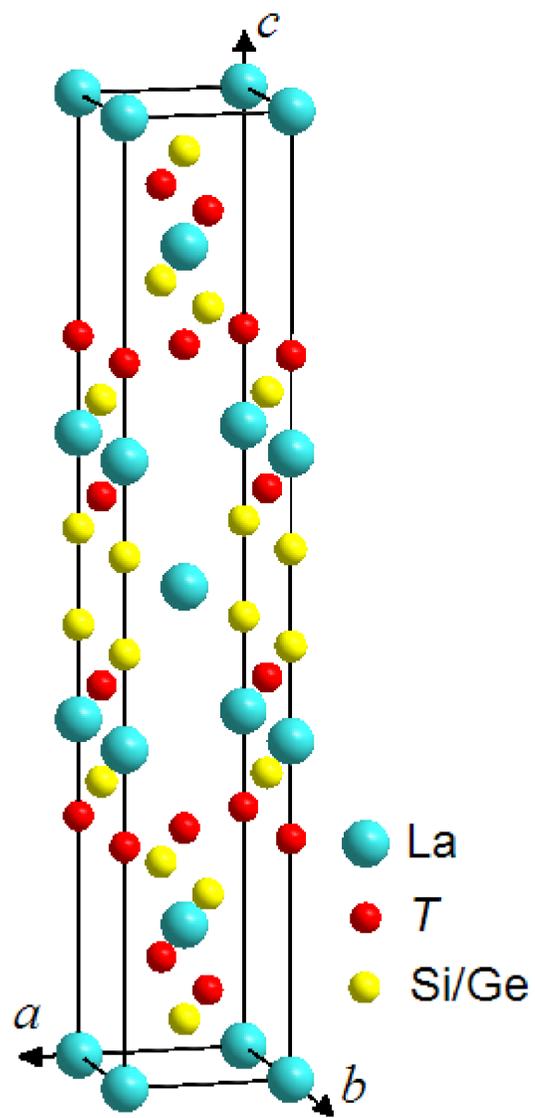

**Fig. 1.** (Color online) Unit cell of $La_3T_4(Si;Ge)_4$ systems ($T$= Ni, Pd, Rh) of the $U_3Ni_4Si_4$-type (*Immm*, no. 71).

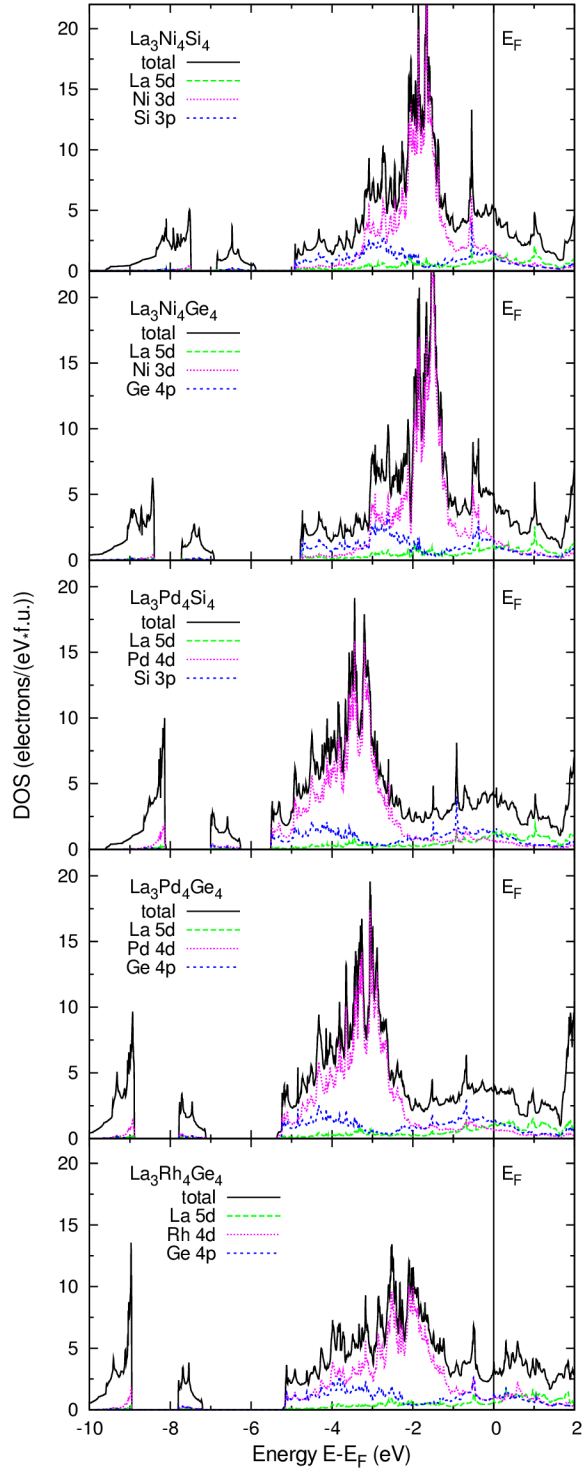

**Fig. 2.** (Color online) Calculated (LDA) total and partial (per electron orbitals of transition-metal, 3d/4d/5d, and other, 3p/4p, atoms) DOSs in La$_3T_4X_4$ for $T$ = Ni, Pd, Rh and $X$ = Si, Ge.

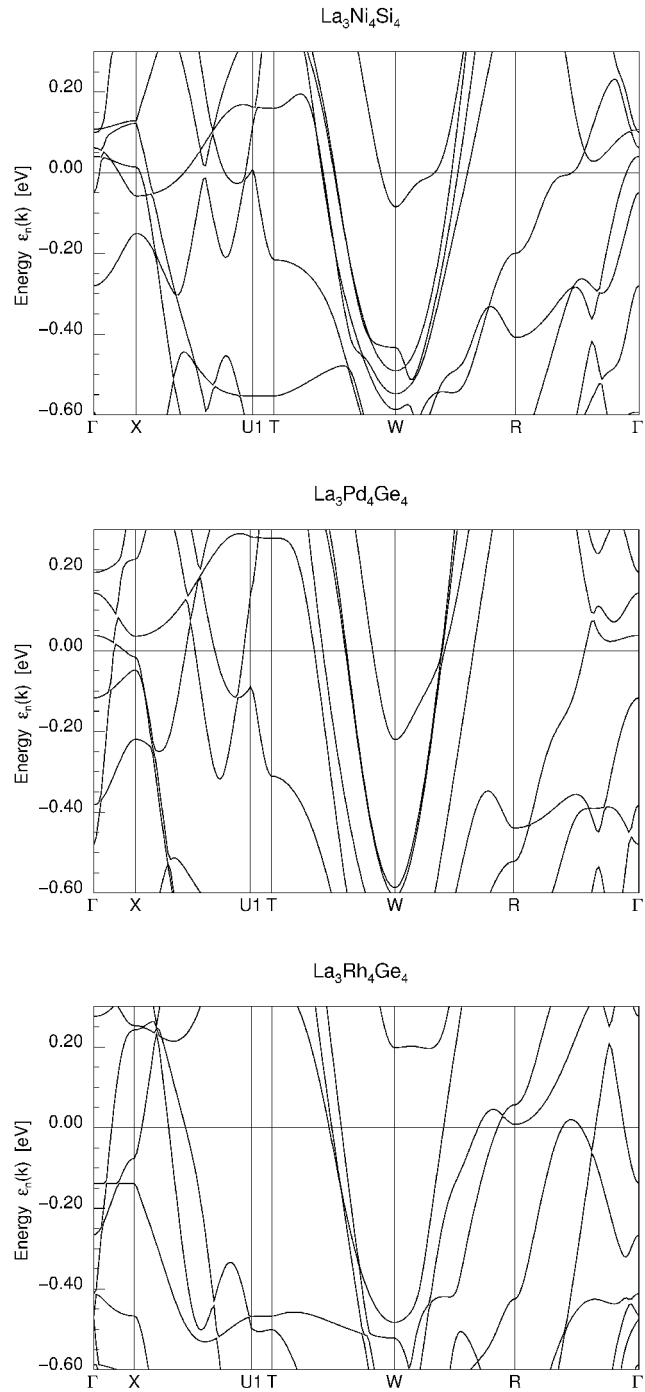

**Fig. 3.** Computed (LDA) band structures of two chosen superconductors, $La_3Ni_4Si_4$ and $La_3Pd_4Ge_4$, compared with that of non-superconducting $La_3Rh_4Ge_4$, displayed in the vicinity of $E_F$.

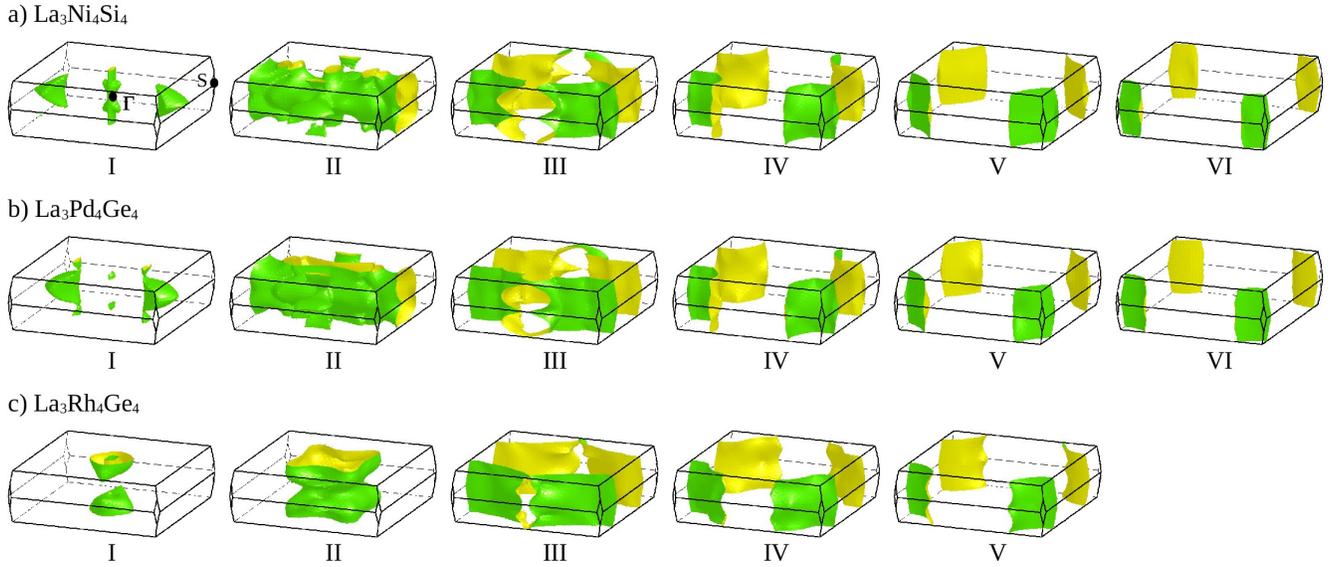

**Fig. 4.** (Color online) Calculated (LDA) FS sheets originating from a few bands (denoted as I-VI), drawn separately within the orthorhombic *Immm*-type BZ boundaries, for superconductors: a) $La_3Ni_4Si_4$ and b) $La_3Pd_4Ge_4$, as well as c) non-superconducting $La_3Rh_4Ge_4$. The FS sheets I-II reveal holelike and III-VI – electronlike characters.

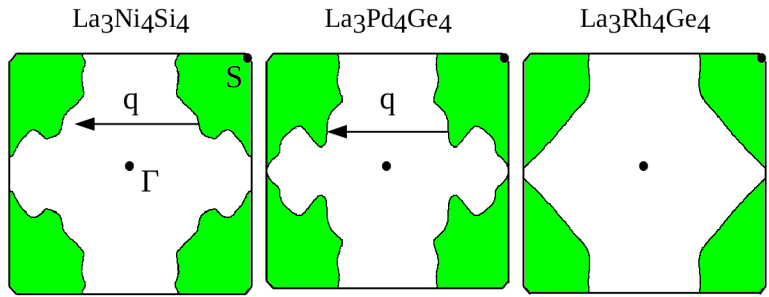

**Fig. 5.** (Color online) Calculated sections of III FS sheets (shown in Fig. 4) through the basal (001) plane, for two superconductors, $La_3Ni_4Si_4$ and $La_3Pd_4Ge_4$, and non-superconducting $La_3Rh_4Ge_4$. The arrow denotes nesting vector $\mathbf{q} = 0.5 \times (2\pi/\mathbf{b})$, being suitable for spanning electron surfaces of the Pd-based superconductors. Green color marks area occupied by electrons.